\documentclass[11pt]{article}

\usepackage{epsfig,amsmath,amssymb,amscd} 
\begin{document}

\begin{center}
{\Large{\bf{Peierls Distortion and Quantum Solitons}}}

\bigskip 

{\large{\bf{Chiara Marletto}}$^{\, 1}$, {\small{and}} {\large{\bf{Mario Rasetti}}$^{\, 2}$}} \\ 
\medskip 
{\small{
$^{1}$Mathematical Institute - Oxford University, \\24-29 St Giles, OX13LB, Oxford (United Kingdom)\\
\smallskip
$^2$ ISI Foundation, Via Alassio 11-C; 10126 Torino (Italy) \\ 
and \\  
Department of Applied Science and Technology, Politecnico di Torino, \\ Corso Duca degli Abruzzi 24; 10129 Torino (Italy) \\ }} 
\end{center}

\medskip
{\footnotesize{
\begin{center} 
{\bf Abstract}
\end{center} 
\noindent Peierls distortion and quantum solitons are two hallmarks of 1-dimensional condensed-matter systems. Here we propose a quantum model for a one-dimensional system of non-linearly interacting electrons and phonons, where the phonons are represented via coherent states. This model permits a unified description of Peierls distortion and quantum solitons. The non-linear electron-phonon interaction and the resulting deformed symmetry of the Hamiltonian are distinctive features of the model, of which that of Su, Schrieffer and Heeger can be regarded as a special case.}} 
\bigskip 

\noindent One-dimensional condensed-matter systems have attracted increasing interest in several branches of physics: not only do they have promising applications in information-processing technologies, but they also play a central role in biological molecules. They consist of linear chains of $'$ions$'$ whose conduction electrons move primarily along the chain axis. Hence, attention can be confined to a single chain. 

\noindent In this paper we shall consider a half-filled 1-d chain, i.e., one consisting of an even number of ions, each of which carries a single conduction electron. This class of systems is particularly interesting because they undergo Peierls distortion: at equilibrium the ions shift from the equally-spaced configuration and assume a dymerised pattern, where the bonds between adjacent ions are alternatively short and long. To date, the best account of this phenomenon is Peierls's theorem, \cite{PEI}, stating that the dimerised configuration minimises the total energy of a 1-d half-filled chain. However, the proofs of the theorem rely on models which represent the ion coordinates as static variational parameters, \cite{FRO},\cite{KELI}. Hence, they do not clarify how Peierls distortion affects the energy spectrum of the whole system, let alone how it contributes to collective, dynamical effects that are also expected to arise.
%If the ion displacement from the equally-spaced configuration is described by a scalar field %depending on the position along the %lattice, the ground-state energy density, as a function of %displacement, is thus expected to have a double-well shape, the zero-%displacement point being %unstable.
%%%%%%%%%%%%%%%%%%%%%%%%%%%%%%%%%%%%%%%%%%%%%%%%%%%%%%%%%%%%%%%%%%%%%
\begin{figure}[htbp]
\centering
\includegraphics[scale=0.6]{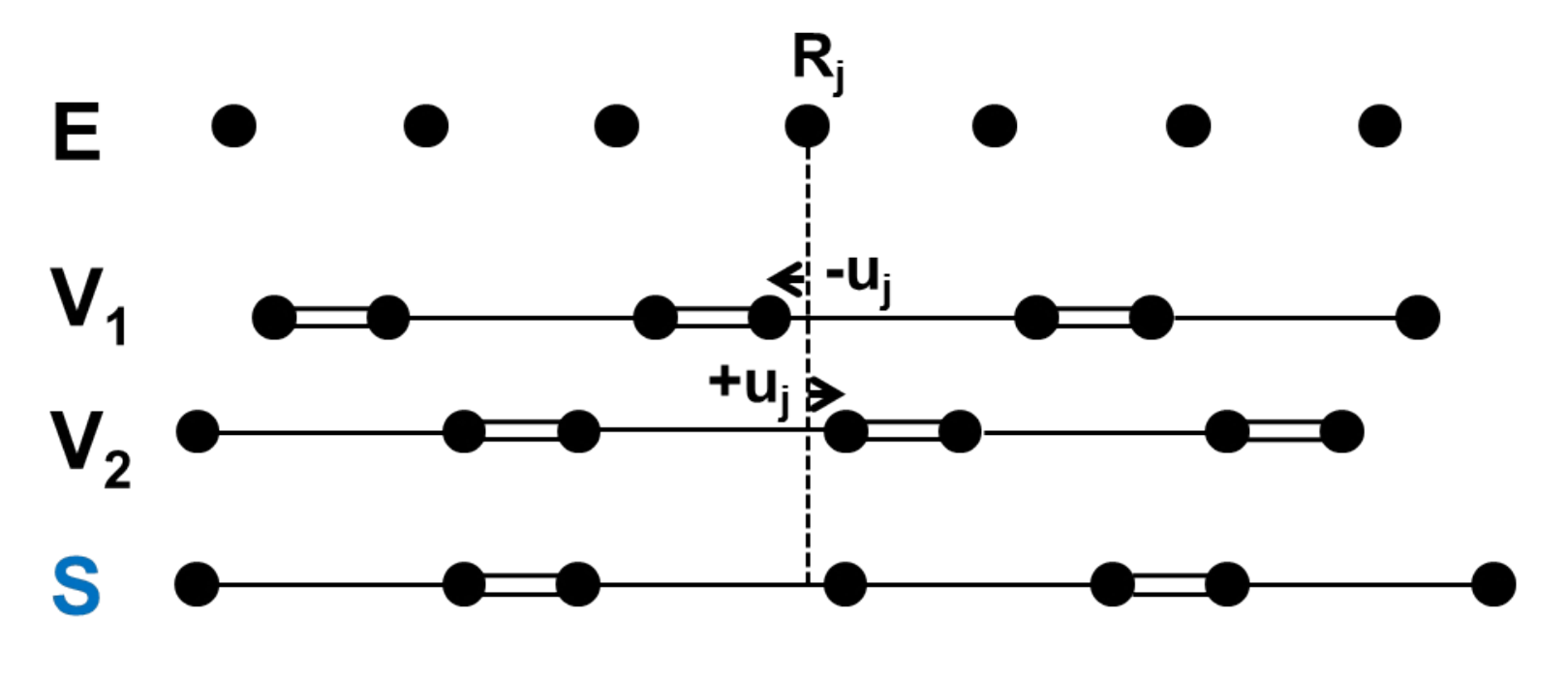} 
\caption{\small{Peierls distortion. The unstable configuration, $E$; the two stable vauca, $V_1$ and $V_2$; a soliton $S$ interpolating between them. }}
\label{s}
\end{figure}
%%%%%%%%%%%%%%%%%%%%%%%%%%%%%%%%%%%%%%%%%%%%%%%%%%%%%%%%%%%%%%%%%%%%%%

\noindent Specifically, since the equally-spaced configuration of identical ions has a reflection symmetry, there are two topologically non-equivalent  ground states ({\sl vacua}) in which Peierls distortion may result, one obtainable from the other by exchanging the position of long and short bonds (fig.\ref{s}).
Hence, there are additional stable states of the system, known as quantum solitons, where the two degenerate vacua coexist (at a given time) and a kink-like domain wall (S in fig.\ref{s}) interpolates between them. Soliton peculiar properties, such as fractional charge eigenvalues, have thus far been described only by phenomenological models \cite{JAC99}, \cite{JAC05}, \cite{JARE}, and so has their connection with Peierls distortion. In fact, the best available model addressing the latter issue $-$ proposed by Su Schrieffer and Heeger (SSH)  for polyacetylene \cite{SUSCHE}$-$ assumes, for phenomenological reasons, the electron-phonon coupling to be linear in the ion displacements, as the expected values of displacements in the distorted ground state are much smaller than the interatomic distance. Overall, we lack of a unified description for the systems sustaining Peierls distortion: the available one  is disconnected and incomplete, since it relies on ad hoc models, valid only in special regimes. 

\noindent In this paper we propose a second-quantized model for a 1-d system supporting Peierls distortion, where the electrons and phonons interact non-linearly. The ion-coordinates are described as semiclassical dynamical variables in coherent states. In this framework we prove Peierls's theorem, showing that Peierls distortion is a direct manifestation of the non-linear electron-phonon interaction, and we show that the system supports a kink-like excitation propagating along the chain at constant energy. 

\noindent {\it The model Hamiltonian and its {\sl quantum} symmetry} $-$  Our model describes a linear chain of $2L$ ions interacting with $2L$ conduction electrons. 
%The state space is ${\cal{H}} =  {\cal{H}}_{el}\otimes {\cal{H}}_{ph}$, where ${\cal{H}}_{el} \sim {{\mathbb{C}}^2}^{\otimes 2L}$ is the overall fermion state space, and ${\cal{H}}_{ph} \sim {\mathfrak{F}}^{\otimes N}$, where ${\mathfrak{F}}$ denotes the (bosonic) single-phonon Fock space. 
%= {\rm span}\left \{|m\rangle\; |\; \hat n|m\rangle=m|m\rangle,\; m\in {\mathbb{N}}\right\}$ ( $\hat n \doteq a^{\dagger}a$, $[a,a^{\dagger}]=1$). 

\noindent The Hamiltonian is: $H = H_{ph} + H_{el-ph}$. Here, $H_{ph} = \frac{1}{2} {\displaystyle \sum_{j=0}^{2L-1} }\bigl ( p_{j}^2 + u_{j}^2 \bigr )$ is the Einstein-phonon Hamiltonian with the ion mass, oscillator frequency and $\hbar $ set equal to 1; $u_{j}$ is the ion displacement from the equilibrium position $R_j$ (see fig. \ref{s}) and $p_{j}$ is the corresponding momentum, $\bigl ( [ u_{j},p_{k} ] = i \hbar \delta_{j,k} \bigr )$. The degrees of freedom in the plane orthogonal to the chain are considered as $'$frozen$\, '$. The electron-phonon Hamiltonian, $$\displaystyle{H_{el-ph} = -\sum_{j=0}^{2L-1} \; t_{j,j+1}  \bigl ( f_{j+1}^{\dagger} f_j + h.c. \bigr )}\;$$ is a tight-binding Hubbard Hamiltonian written  in terms of fermionic creation and annihilation operators $f_j^{\dagger}$, $f_j$ $\bigl ( \{ f_j , f_{\ell}^{\dagger} \} = \delta_{j,\ell}$, $\{ f_j , f_{\ell} \}=0 \bigr )$.  The index $j$ includes both electron position and spin, but the latter is irrelevant in the present context and will not be explicitly written.  The $'$amplitude$\;'$ for an electron $'$hopping$\, '$ from site $j$ to site $j+1$ is the operator  $t_{j,j+1}$, depending on the phonon degrees of freedom. It includes, as argued in \cite{MORA}, the electron-phonon coupling:
\begin{eqnarray}
{{t}}_{ j, j+1} = T \, \exp \left ( \zeta \left (u_{j+1} - u_{j} \right)\right ) \, \exp \left ( \kappa \left ( p_{j+1} - p_{j} \right ) \right ) \quad \zeta,\;\kappa \in \mathbb{R},\; T\in \mathbb{C}, \nonumber
\end{eqnarray}
where $\zeta$ and $\kappa$ depend on the form of the Wannier functions out of the ion core \cite{CAMA}. Even though no explicit phonon-phonon interaction is considered, the phonons are indirectly coupled via $H_{el-ph}$. The SSH Hamiltonian can be recovered by setting $\kappa=0$ and approximating $t_{j,j+1}$ to first order in $\zeta$.

\noindent The Hamiltonian $H_{el-ph}$ has a local dynamical symmetry and a global symmetry, both associated with the Hopf {\sl quantum} algebra ${\cal U}_q(su(2))$, \cite{MORA}. The generators of this algebra $\{K^{(\pm)}, K^{(3)}\}$ close the $q$-deformed commutation relations, \cite{FU}, $[K^{(3)},K^{(\pm)}]=\pm K^{(\pm)}$, $[K^{(+)},K^{(-)}]= [K^{(3)}]_q$, where $[A]_q\doteq \frac{q^{A}-q^{-A}}{q-q^{-1}}$ and $q$ is the deformation parameter (which can be assumed to be real). The algebra ${\cal U}_q(su(2))$ belongs to the universal envelope of $su(2)$ and reduces to the latter for $q\longrightarrow 1$. As discussed in \cite{MORA}, the generators of the local dynamical symmetry contain both fermionic and bosonic operators, and so do the generators of the global symmetry (defined via co-producting the local operators). Also, $q$ depends on the physical constants of the model, in such a way that  these symmetries reduce to a non-deformed $su(2)$ symmetry (i.e., $q\longrightarrow 1$) if  $\zeta \rightarrow 0$ or $\kappa \rightarrow 0$. 
Since both $\zeta$ and $\kappa$ are non-negligible in the systems supporting Peierls distortion, we expect that the {\sl quantum} symmetry will play a central role in describing this phenomenon. 

 \noindent {\it {Staggering in phonon coherent states}} $-$ In view of the different time-scales of the phonon 
and the electron dynamics we adopt for the phonon operators a semiclassical, dynamical description in terms of Glauber coherent states \cite{PER}.  The global coherent state of the ions is $\displaystyle{| \mathcal Z \rangle \equiv \bigotimes_{j} |
z_j \rangle}$, $1\leq j\leq 2L-1$, where $\displaystyle{| z_j \rangle = {\rm{e}}^{- \frac{1}{2} |z_j|^2} \, {\rm{e}}^{z_j 
a_j^{\dagger}} |0\rangle_j}$, $a_j$ is the $j$-th single-phonon creation operator, $ |0\rangle_j$ is the vacuum state such that  $a_j |0\rangle_j=0$ and $z_j \; \in\; {\mathbb{C}}$. Here, ${\rm{Re}} z_j$ and ${\rm{Im}} z_j $ represent respectively the 
expected values in coherent states of the $j$-th ion
displacement and momentum. 

\noindent Inspired by \cite{SUSCHE}, we first explore the possibility of a staggered ground state  by setting $z_j = (-)^j z$, where $z$ is a variational parameter to be found minimising the ground-state energy ($z=0$ corresponds to the non-dimerised configuration, $E$ in fig.\ref{s}). 
The staggering condition is well defined, as it involves the {\sl expected values} of the ion positions and momenta in coherent states; in constrast, in \cite{SUSCHE} the condition is imposed on the ion position {\sl operators}, ignoring the effect on the conjugate momenta, which may lead to a difficulty with Heisenberg principle. Besides, global momentum conservation implies, in our semiclassical picture, that the ion momenta are staggered too.
We also require $T \doteq t\exp(-i\hbar \zeta \kappa)$, with $t\;\in \mathbb{R}$, so that $\langle \mathcal Z | t_{j,j+1} |\mathcal Z \rangle \in \mathbb{R}$ (time-reversal symmetry). 
 
\noindent {\it The averaged Hamiltonian and its symmetry} $-$ Representing the Hamiltonian in coherent states we find: ${\hat{H}} \equiv \langle \mathcal Z | H | \mathcal Z  \rangle$ $= {\hat{H}}_{ph} + {\hat{H}}_{el-ph}$ $\equiv \langle \mathcal Z | H_{ph} | \mathcal Z \rangle$ $+ \langle \mathcal Z | H_{el-ph} | \mathcal Z \rangle$ where ${\hat{H}}_{ph}$ $= 2 L \, \left( 4( {\rm Re}z )^2 + ( {\rm Im}z )^2 + \frac{3}{4} \right)$ (a $c$-number) and
\begin{eqnarray} 
{\hat{H}}_{el-ph} = -g \sum_{j=0}^{2L-1} \left ( \cosh({\mathfrak{z}}) - (-)^j \sinh({\mathfrak{z}}) \right ) \, \left ( f_{j+1}^{\dagger} f_j + \rm{h.c.} \right ) \; , \nonumber 
\end{eqnarray} 

\noindent where $'$state location$\, '$ ${\mathfrak{z}}\doteq 2\sqrt{ 2 }\left ({\rm Re}(\zeta z) + {\rm Im}\; (\kappa  z)\right )$ and effective coupling $g\doteq t \exp \left (  \zeta^2 + \kappa^2  \right)$ have been introduced.

\noindent Fourier-transforming the $\{f_j\}$ into the standard particle-hole fermionic operators $\{c_k, v_k\}$, $0 \leq k \leq L-1$, we have  $\displaystyle{{\hat{H}}_{el-ph} = \sum_{k=0}^{L-1} H_k }$, with 

\begin{equation}
H_k \doteq  -2\epsilon J_{3} -\delta \left [J_{+} + J_{-}\right ] \; , \label{qcomp}
\end{equation}

\noindent where $\epsilon\equiv\epsilon (k,z)\doteq g\; \cosh({\mathfrak{z}})\cos\left(\frac{\pi}{L}k\right)$, $\delta\equiv\delta (k,z)\doteq g\; \sinh ({\mathfrak{z}}) \sin \left(\frac{\pi}{L}k \right 
)$ and $J_{+} \doteq v_{k}^{\dagger} c_k$, $J_{-}\doteq J_{+}^{\dagger}$, $J_{3} \doteq \frac{1}{2} \left ( n_{k}^{(v)} - n_{k}^{(c)} \right )$ (dropping the mode-index $k$ for simplicity). 

\noindent The operators $J_{\alpha }\, , \, \alpha \in \{ + , - , 3\}$, close an $su(2)$ algebra in the spin-$\frac{1}{2}$ representation, 
${\cal D}_{\frac{1}{2}}$. ${\hat{H}}_{el-ph}$ has therefore the dynamical symmetry described by the algebra ${\cal A } = \bigoplus_k su(2)_{(k)}$, whereas the original Hamiltonian had both a global symmetry and a local dynamical symmetry associated with the $q-$deformed algebra ${\cal U}_q(su(2))$. Indeed, the generators of the quantum symmetry \cite{MORA}, when averaged in coherent states, lose their dependence on the phonon operators and reduce to the generators of $su(2)$. Since the quantum symmetry is induced by the electron-phonon interaction and the latter is central to the description of systems supporting Peierls distortion, we shall now define a procedure to restore it.
% Moreover, in $| \mathcal Z \rangle$ the ion degrees of freedom are decoupled from each other, unlike what happens in the co-product.

\noindent {\it Restoring the {\sl  quantum} symmetry} $-$ To this end, it is not convenient to refer to ${\cal U}_q(su(2))$, as it is not a proper algebra and does not provide the group operation to diagonalise the Hamiltonian. Instead, we define a proper algebraic structure, ${\cal A}_q$, having more appealing properties. Specifically, ${\cal A}_q$ is the three-dimensional submodule of ${\cal U}_q(su(2))$ closed with respect to the deformed adjoint action, $\bigl [ \; ,\, \bigr ]_q:$  $\;\bigl [ K^{(\pm )} , f \bigr ]_q= K^{(\pm )} f  q^{K^{(3)}} - q^{K^{(3)} \mp 1} f K^{(\pm )}$, $\displaystyle{\bigl [ K^{(3)} , g \bigr ]_q = K^{(3)} g - g K^{(3)}}$ and $\bigl [ f g , h \bigr ]_q = \bigl [ f , \bigl [ g , h \, \bigr ]_q \;\bigr ]_q$, $\forall f, g, h \in {\cal{U}}_q\bigl ( su(2) \bigr )$. Resorting to the deformed adjoint action is necessary since there is no three dimensional submodule closed with respect to the deformed commutation relations in ${\cal U}_q(su(2))$.  ${\cal A}_q$ is generated by the operators
\begin{equation}
J_{\pm}^{(q)} \doteq q^{w}\sqrt{\xi_q}\, q^{-K^{(3)}\pm \frac{1}{2}}K^{(\pm )} \; ; \; \;
J_3^{(q)} \doteq q^{2w}\frac{ \xi_q }{2}\left ( q K^{(+)} K^{(-)} - q^{-1} K^{(-)} K^{(+)} \right ) \; , \nonumber %\label{Lq}
\end{equation}   
\noindent where $K^{(\pm )},\; K^{(3)}$ are the generators of ${\cal U}_q(su(2))$, $\xi_q\doteq 2q^{-w}{(q+q^{-1})}^{-1}$ and $q,w \in \mathbb{R}$.  Also, ${J_{+}^{(q)}}^{\dagger}={J_{-}^{(q)}}$. Notice that for $q\to 1$ ${\cal A}_q$ coincides with $su(2)$, as $\bigl [ \; ,\, \bigr ]_1\equiv \bigl [ \; ,\, \bigr ]$.
%
%\medskip 
%\noindent Operators \eqref{Lq} satisfy:
%\begin{eqnarray}
%\bigl [ J_+^{(q)} , J_-^{(q)} \, \bigr ]_q &=& 2 J_3^{(q)} = - \bigl [ J_-^{(q)} , J_+^{(q)} \, \bigr ]_q \; , \nonumber \\
%\bigl [ J_3^{(q)} , J_{\pm}^{(q)} \, \bigr ]_q &=& \pm \, q^{w\pm 1} J_{\pm}^{(q)} = - q^{\mp 2} \bigl [ J_{\pm}^{(q)} , J_3^{(q)} \, \bigr ]_q \; , \nonumber \\
%\bigl [ J_3^{(q)} , J_3^{(q)} \, \bigr ]_q  &=& q^{w}\bigl ( q - q^{-1} \bigr ) \, J_3^{(q)} \; , \; \bigl [ J_{\pm}^{(q)} , J_{\pm}^{(q)} \, \bigr ]_q = 0 \; ,\label{qcomm}
%\end{eqnarray} 
%
%
%%\footnote{However, it is not a Lie algebra since $\bigl [  \; ,  \, \bigr ]_q$  does not satisfy the antisymmetry nor the Jacobi identity.}  

\noindent In order to restore the quantum symmetry we simply replace in $H_k$ each operator $J_{\alpha}$ with $J_{\alpha}^{(q)}$, for every $k$, i.e., $H_k$ with
$$
H_k^{(q)} =  -2\epsilon J_3^{(q)} -\delta\left (J_+^{(q)} + J_-^{(q)}\right ) \;.  
$$

\noindent The numbers $q$ and $w$ therefore become parameters of the model. This Hamiltonian, formally identical with \eqref{qcomp}, is endowed with a quantum dynamical symmetry which retains memory of the of the original Hamiltonian symmetry.  Indeed, as a side remark, a physical interpretation of the quantum symmetry restoration may be provided by writing the generators of ${\cal A}_q$ in terms of $'' q$-deformed$\, ''$ fermionic-like operators, ${\mathfrak{f}} , {\mathfrak{f}}^{\dagger}$, obeying deformed Clifford anticommutation relations: $\bigl \{ {\mathfrak{f}} \, , {\mathfrak{f}}^{\dagger} \bigr \} = \frac{\sinh (2Q)}{2 Q}$, ${\mathfrak{f}}^2 = {{\mathfrak{f}}^{\dagger}}^2 = - \frac{1}{2} \, Q \left ( \frac{\sinh Q}{Q} \right )^2$, $Q\sim \log(q)$. These operators represent $''$dressed $\, ''$ electrons, retaining some memory of the interaction with phonons.

%One might as well extend this symmetry to the global level using the co-algebra, and determining then the free parameters $q$ and $w$ in such a way that the resulting global algebra commutes with $H^{(q)}$. We shall not address this point here, because it plays no role in the physics of Peierls instability.
\noindent  In the $D_{\frac{1}{2}}$ representation, ${\cal A}_q \sim u(2)=u(1)\oplus su(2)$ and the eigenvalues of  $H_k^{(q)}$ are:
\begin{equation}
\Lambda_{\pm}= - \frac{1}{2}\epsilon (q-q^{-1})q^{2w}\xi_q\mp \sqrt{q^{2w}\epsilon^2+\xi_q\delta^2}\equiv \Lambda_{\pm}{(k,z)}\; . \nonumber 
\end{equation}
(which, in the limit $q\longrightarrow 1$, tend to the eigenvalues of $H_k$.)
Since in  $D_{\frac{1}{2}}$ the co-product is primitive (${\cal A}_q \sim u(2)$) the global symmetry is automatically restored  by setting $\displaystyle{\hat H_{el-ph}=\sum_k H_k^{(q)}}$. The energy spectrum of the system is therefore the sum over the modes $k$ of the eigenvalues $\Lambda_{\pm}$.

\noindent {\it Proving Peierls's theorem} $-$ The ground state energy density has the form 
$\displaystyle{{\cal E}_q(z)=\frac{1}{L}\sum_{k=0}^{L-1} \Lambda_{+}(k,z)}$ (taking into account the spin degeneracy by a factor $2$). In the limit $L \gg 1$, 
\begin{equation}
{\cal E}_q(z) = -p_q(z) \int_{0}^{\frac{\pi}{2}}{\rm d}{k}\sqrt{1- m_q \;\sin^{2}(k)}\, \nonumber 
\end{equation}
with $\displaystyle{p_q(z)\doteq  \frac{2}{\pi} \, g q^{w} \, \cosh({\mathfrak{z}})}$, $\displaystyle{m_q \doteq (1- y_q^2)}$, $\displaystyle{y_q\doteq \sqrt{\xi_q} \, \tanh({\mathfrak{z}})}$. 

%%%%%%%%%%%%%%%%%%%%%%%%%%%%%%%%%%%%%%%%%%%%%%%%%%%%%%%%%%%%%%%%%%%%%

\begin{figure}[htbp]
\centering
\includegraphics[height=5cm, width=6cm]{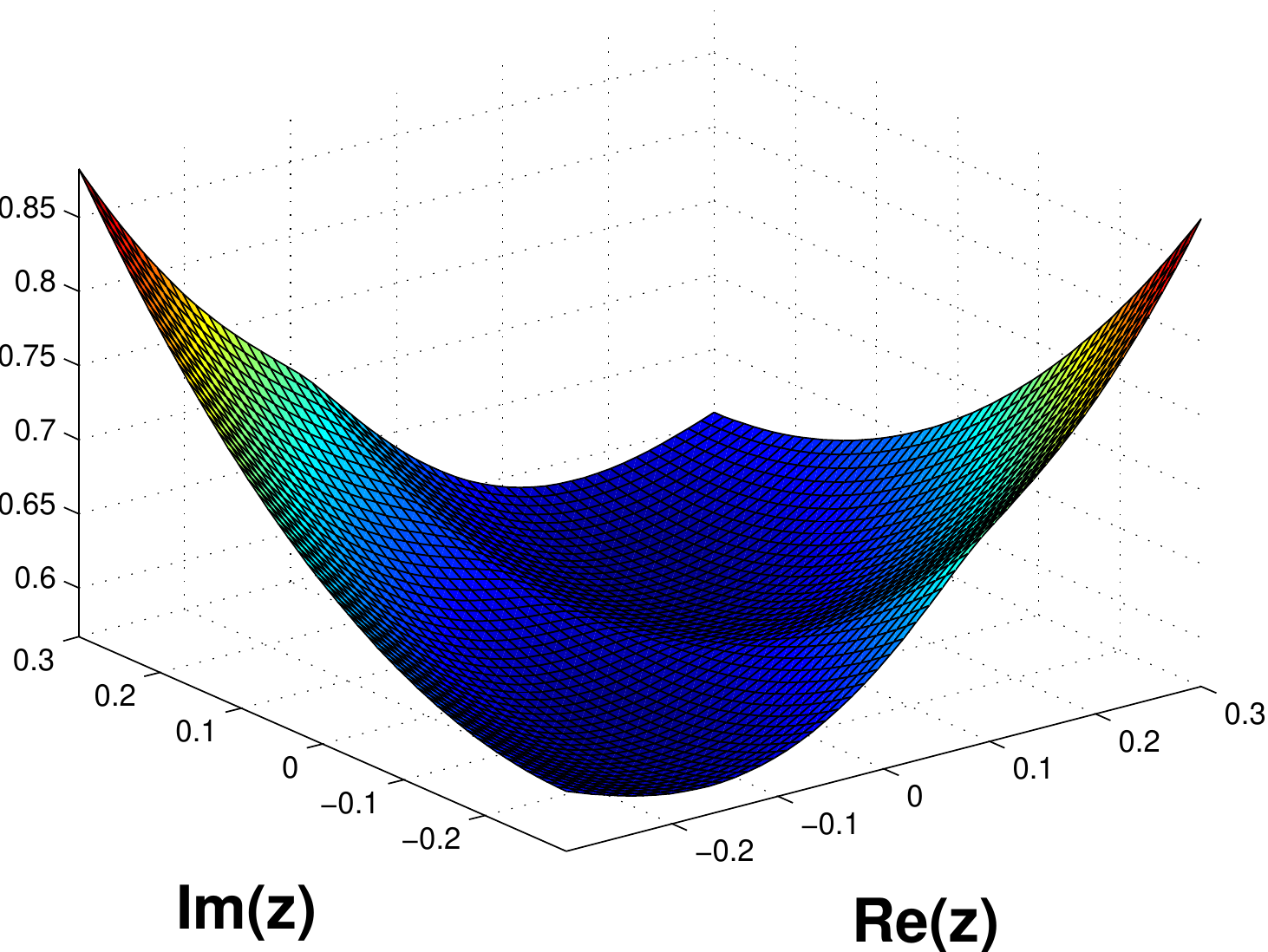}
\quad 
\includegraphics[height=5cm, width=6cm]{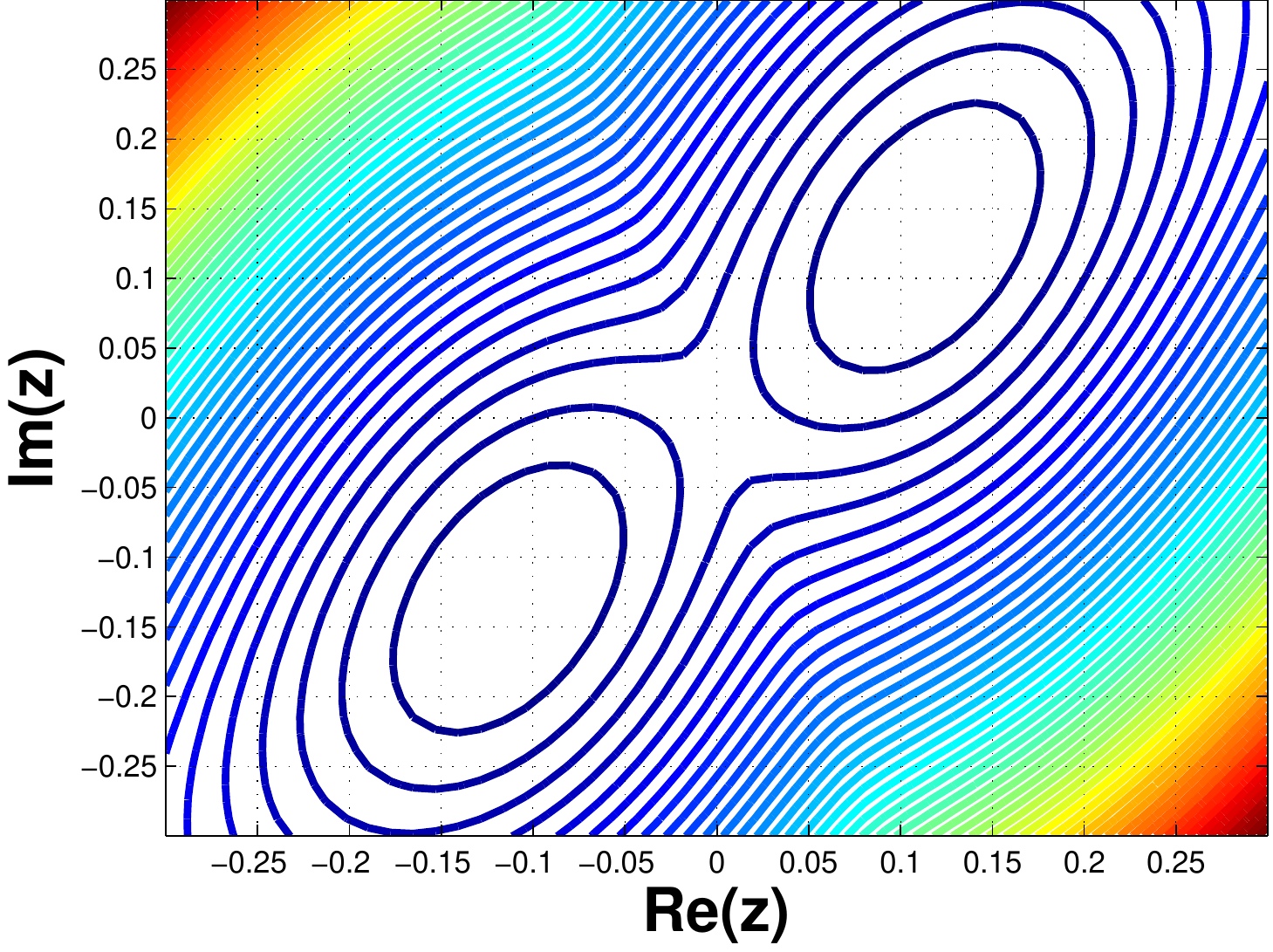} 
\caption{\small{The ground state energy density when $1<\xi_q<2$ and $q=1.5$.}}
\label{1}
\end{figure}

%%%%%%%%%%%%%%%%%%%%%%%%%%%%%%%%%%%%%%%%%%%%%%%%%%%%%%%%%%%%%%%%%%%%%%

\noindent The integral in ${\cal E}_q(z)$ for $0 \leq \xi_q \leq1$, i.e., $0 \leq m_q \leq 1$ is an elliptic integral of the second kind; whereas for $1<\xi_q\leq 2$, $-1\leq m_q\leq1$, and in general for $\xi_q>2$, it is the hypergeometric function ${_2F_{1}}(\frac{1}{2},-\frac{1}{2},1; m_q)$. The integral is real and converges if $|\mathfrak{z}|\leq\zeta_m$, where $\zeta_m$ is the value for which $|m_q|=1$ \cite{ABST}. This allows one to perform a detailed study of the total energy density of the ground state, ${\cal E}_{G.S.}= {\cal E}_{ph}(z)+{\cal E}_q(z)$.

\noindent  For appropriate values of $q$ and $\xi_q$ (i.e., $w$) ${\cal E}_{G.S.}$ exhibits (see fig.(\ref{1})) a saddle point in ${\rm Re}z=0={\rm Im}z$ (corresponding to the equally-spaced configuration) and two degenerate minima in $\pm({\rm Re}z, {\rm Im}z)\neq 0$, corresponding to the two degenerate ground states induced by Peierls distortion. This shows that the dimerized configuration minimizes the total energy and proves Peierls's theorem, as promised.  For $q\to 1$ the total energy has only one maximum with zero second derivative, describing a marginally stable equilibrium for $z=0$. This limit includes, for $\kappa=0$ and ${\mathfrak{z}} \ll 1$, the SSH model (linear in ${\rm Re}(z)$).  This shows that the deformed symmetry is decisive to describe Peierls effect.  

\noindent {\it Dynamical equations in the ground state} $-$ In the coherent-state formalism $x\doteq2{\rm Re}z$ and $p\doteq 2{\rm Im}z$ are dynamical variables specifying the ion canonical coordinates. ${\cal E}_{G.S.}(x,p)$ can then be considered as the (classical) Hamiltonian describing the dynamics of the system ground state in phase space ${\mathbb{C}}^2$ \cite{DAMA}. The motion of the representative point $(x(t), p(t))$ describes the collective evolution of the phonons interacting with the electrons in the ground state. The corresponding equations of motion along the line $x=p$ is  (see fig.\ref{1}) represent a non-linearly damped, non-linearly driven oscillator: \begin{equation}
\ddot x = (x- {\cal P})\left (1-{\cal P}_x\right)-\dot x{\cal P}_x\;, \nonumber
\end{equation}
where ${\cal P}_x\doteq \frac{\partial {\cal P}}{\partial x}$ and

$$
{\cal{P}} (x,p) \doteq \frac{4\sqrt{2}g}{\pi}\frac{\sinh\left(\sqrt{2}(x+p)\right)}{\xi_q (q+q^{-1})} \left ( E(m_q)- \xi_q\frac{E(m_q)-F(m_q)}{m_q[\cosh(\sqrt{2}(x+b))]^2} \right ) \;  
$$
with $ E(m_q)\doteq{_2 F_{1}}(\frac{1}{2},-\frac{1}{2},1; m_q)$ and $ F(m_q)\doteq {_2 F_{1}}(\frac{1}{2},\frac{1}{2},1; m_q)$. The presence of a damping factor shows that the Peierls-distorted ground state is robust against perturbations. Besides, this non-linear equation strongly suggests the existence of soliton excitations.

\noindent {\it The existence of solitons } $-$ In analogy with the procedure used to prove Peierls's theorem, we explore the possibility of an excited kink-like state at site $n$. To do so, we adopt as a trial description of the phonon coherent state 
 $\displaystyle{| \mathcal Z_n \rangle \doteq \bigotimes_{j=0}^{n} | z_j \rangle \otimes \bigotimes_{\ell = n+1}^{N-1} | z_{\ell} \rangle}$, where the deformed staggering condition
$$
z_j = (-)^j z \; , \; {\rm{for}} \quad 0 \leq j \leq n \quad , \quad   
z_{\ell} = (-)^{\ell +1} z \; , \; {\rm{for}} \quad n+1 \leq \ell \leq N-1 \; ,
$$
represents the presence of a kink at site $n$. The averaged Hamiltonian ${{\hat{H}}\,'}_{n} \doteq \langle \mathcal Z_n | H_{el-ph} | \mathcal Z_n \rangle$ is $$
{{\hat{H}}\,'}_{n} = \sum_{j=1}^{N-1} \omega_j f_{j+1}^{\dagger}f_{j} + \omega_n f_{n+1}^{\dagger}f_n - 2s \sum_{j=n+1}^{N-1}(-)^j f_{j+1}^{\dagger} f_{j} + {\rm{h.c.}} \; ,
$$ 
where $\omega_{\ell} \doteq g - \bigl ( c + (-)^{\ell} s \bigr )$, $s=g\sinh({\mathfrak{z}})$, $c=g\cosh({\mathfrak{z}})$, and the energy spectrum can be obtained by repeating the diagonalisation procedure via pseudo-fermionic operators. To prove the existence of solitons we shall argue that the dynamics generated by ${{\hat{H}}\,'}_{n}$, for appropriate initial states, induce the kink to move spontaneously from site $n$ to site $n+1$. 

\noindent  The state at time $t$, consisting of the fermionic (pseudo-spin) and bosonic components, can be written as $| \psi (t) \rangle = {\rm{e}}^{- \frac{1}{2} | z (t) |^2} \, {\rm{e}}^{z (t) a^{\dagger}} \, | 0 \rangle \otimes{\rm{e}}^{- i t H_{n}' ( z (t) )} \, | s(0) \rangle $, where $z(t)$ satisfies the canonical equations ${\dot{z}} = \frac{\partial {\cal{E}}}{\partial {\bar{z}}} \; , \; {\dot{\bar{z}}} = - \frac{\partial {\cal{E}}}{\partial {z}} \; $ and ${\cal{E}} (z, n)$ is the lowest eigenvalue of $\hat H_n'$, representing the first excited state of the system. If the time $\delta t$ of the kink motion from site $n$ to $n+1$ is very small, one has $| \psi (t + \delta t) \rangle = \left [ 1 + \delta t \left ( {\dot{z}} \bigl ( a^{\dagger} - {\bar{z}} \bigr ) \otimes {\mathbb{I}} - i {\mathbb{I}} \otimes (\hat H'_n + {\dot{z}} \, \frac{\partial H'_n}{\partial z}) \right ) \right ] \, | \psi (t) \rangle$ to first order in $\delta t$ and $ \;\,\;\frac{\partial \hat H_n'}{\partial z}\simeq\frac{\partial n}{\partial z}{\cal{D}}_H^{(n+1,n)}$, where $$ {\cal{D}}^{(n+1,n)} \doteq {{\hat{H}}\,'}_{n+1} - {{\hat{H}}\,'}_{n} = \omega_n \bigl ( f_{n+2}^{\dagger} f_{n+1} -   f_{n+1}^{\dagger} f_{n} + {\rm{h.c.}} \bigr )\;. \; $$
This operator has a doubly-degenerate $0$-eigenvalue, whose corresponding eigenspace, $V_0$, is spanned by the states $\cdots \, \otimes | n_n \rangle \otimes | n_{n+1} \rangle \otimes | n_{n+2} \rangle \otimes \, \cdots$, with $n_j = 0 , 1$, and (conserved) total number $n_n + n_{n+1} + n_{n+2} = 1$. 

\noindent Suppose now that $|{\psi(t)}\rangle$ is a superposition of $\{\phi_k^{(n)}\}$, the projections of the eigenstates of ${{\hat{H}}\,'}_{n}$ onto $V_0$, describing the kink at site $n$. Considering a time increment $\delta t$ such that the kink moves at site $n+1$ while the coherent state representative point $z$ changes to $z + \delta z = z + {\dot{z}} \delta t$,  ${\dot{z}}\, \frac{\partial n}{\partial z} = 1$, and using the defining properties of $V_0$, $|\psi(t+\delta t)\rangle$ turns out to be a superposition of $\{\phi_k^{(n+1)}\}$, describing the kink at site $n+1$. Hence, for appropriate initial conditions, the kink moves spontaneously from site $n$ to $n+1$, for all $n$. This proves that the system supports a kink-like excitation propagating along the chain at constant energy, once more as promised.

%%%%%%%%%%%%%%%%%%%%%%%%%%%%%%%%%%%%%%%%%%%%%

\noindent {\it In conclusion} $-$ We have proposed a quantum model for a $1$-d chain of electrons and phonons, providing a self-consistent description of Peierls distortion and showing that the system is able to sustain soliton excitations. Key features of our approach are the non-linear coupling beteween electrons and phonons, which generalises the SSH model; the description of the phonon degrees of freedom in coherent-states, which permits a dynamical picture of the phonons; and the subsequent restoring of the original quantum dynamical symmetry. Further insight  may be gained in the future via a thorough analysis of the ground state and of the soliton dynamics. This may open new perspectives  for the application of solitons in one-dimensional quantum systems. For instance, solitons may be used as means of transferring quantum information. Work is in progress along these lines.  

\noindent {\it Acknowledgements $-$}  CM is supported by EPSCR and Istituto Superiore Mario Boella.


\begin{thebibliography}{99}  
\bibitem{PEI} R. Peierls, {\it Quantum Theory of Solids}, Clarendon, Oxford, (1955).
\bibitem{FRO} H. Fr\"olich, {\sl Proc. Roy. Soc.} {\sl A} {\bf 223}, 296 (1954).
\bibitem{KELI} T. Kennedy, E. Lieb, {\sl Phys. Rev. Lett.} {\bf 59}, 1309 (1987).
\bibitem{JAC99} R. Jackiw, {\it Effects of Dirac's Energy sea on quantum numbers}, Dirac Prize Lecture, Trieste, Italy, (1999).
\bibitem{JAC05} R. Jackiw, {\it Topology in Physics}, TOP 2005 Symposium, Sapporo, Japan, (2005).
\bibitem{JARE} R. Jackiw, C. Rebbi, {\sl Phys. Rev. Lett. D} {\bf 13}, 3398 (1976).
\bibitem{SUSCHE} W. P. Su, J. R. Schrieffer, A. J. Heeger, {\sl Phys. Rev. Lett.} {\bf 42}, 1968 (1979) and {\sl Phys. Rev. B} {\bf 22}, 2099 (1980).
%\bibitem{YULU} {\it Solitons and Polarons in Conducting Polymers}, Y. Lu (Ed.), World Scientific Publ. Co., Singapore, 1988 
%\bibitem{MASU} N. Manton, P. Sutcliffe, {\it Topological Solitons}, Cambridge University Press, Cambridge, 2004
\bibitem{MORA} A. Montorsi, M. Rasetti, Phys. Rev. Lett. {\bf 72}, 1730 (1994) .
\bibitem{CAMA} D. K. Campbell, and S. Mazumdar, in Conjugated Conducting Polymers, Springer-Verlag Series in Solid State Science Vol. 102, edited by D. Baer- H. Kiess, Springer-Verlag, Berlin, (1992).
\bibitem{FU} J. Fuchs, {\sl Affine Lie Algebras and Quantum Groups: An Introduction, With Applications in Conformal Field Theory}, Cambridge University Press, Cambridge (1995).
\bibitem{PER} A. Perelomov, {\it Generalized Coherent States and Their Applications}, Springer-Verlag, Berlin Heidelberg, 1986.  
\bibitem{ABST} {\it Handbook of Mathematical Functions}, M. Abramowitz, I. A. Stegun (Ed.), Dover Publications, Inc., New York, 1965.
\bibitem{DAMA} M. G. Rasetti, G. M. D'Ariano, A. Montorsi, {\sl Phys. Lett.} {\bf 10 7A}, 291 (1985).
\end{thebibliography}
\end{document}